\documentclass[structabstract]{aa}
  \usepackage{txfonts}
   \usepackage{graphicx}
    \usepackage{natbib}
     \usepackage{longtable}
      \usepackage{subeqnarray}
       \usepackage{cases}
 \bibpunct{(}{)}{;}{a}{}{,}

\begin{document}

\title {Filament L1482 in the California molecular cloud
\thanks{Appendices are only available in electronic forms
at the website http://www.aanda.org.}}

\author{D. L. Li\inst{1,2}
 \and J. Esimbek\inst{1,3}
  \and J. J. Zhou\inst{1,3}
  \and Y.-Q. Lou\inst{4}
   \and G. Wu\inst{1,2,3}
    \and X. D. Tang\inst{1,2}
     \and Y. X. He\inst{1,2}}


\institute{Xinjiang Astronomical Observatory, Chinese
 Academy of Sciences, Urumqi 830011, P. R. China\\
 \email{lidalei@xao.ac.cn}
  \and University of the Chinese Academy of Sciences,
  Beijing 100080, P. R. China\\
   \and Key Laboratory of Radio Astronomy, Chinese
   Academy of Sciences, Urumqi 830011, P. R. China\\
   \and Department of Physics and Tsinghua Center for Astrophysics (THCA), Tsinghua University, Beijing 100084, P. R. China \\
   \email{louyq@tsinghua.edu.cn} }

\date{Received 25 November 2013/ Accepted 28 May 2014 }

\abstract
{}
{  The process of gravitational fragmentation in the
 L1482 molecular filament of the California molecular
 cloud is studied by combining several complementary
 observations and physical estimates.
We investigate the kinematic and dynamical states of this
 molecular filament and physical properties of several
 dozens of dense molecular clumps embedded therein. }
{ We present and compare molecular line emission
 observations of the $J=2-1$ and $J=3-2$ transitions
 of $^{12}$CO in this molecular complex, using the
 K\"olner Observatorium f\"ur Sub-Millimeter Astronomie
 (KOSMA) 3-meter telescope.
These observations are complemented with archival data
 observations and analyses of the $^{13}$CO $J=1-0$
 emission obtained at the Purple Mountain Observatory
 (PMO) 13.7-meter radio telescope at Delingha Station in
 QingHai Province of west China, as well as infrared emission
 maps from the Herschel Space Telescope online archive, obtained
 with the SPIRE and PACS cameras.
Comparison of these complementary datasets allows for a comprehensive
 multiwavelength analysis of the L1482 molecular filament.}
{We have identified 23 clumps along the molecular filament
 L1482 in the California molecular cloud.
 For these molecular clumps, we were able to estimate
 column and number densities, masses, and radii.
The masses of these clumps range from  $\sim 6.8$
 to $62.8 M_\odot$ with an average value of
 $ 24.7_{-16.2}^{+31.1}M_\odot$.
Eleven of the identified molecular clumps appear
 to be associated with protostars  and are thus
 referred to as protostellar clumps.
 Protostellar clumps and the remaining starless clumps
 of our sample appear to have similar temperatures  and linewidths,
 yet  on average, the protostellar clumps
 appear to be slightly more massive than the latter.
 All these molecular clumps show supersonic nonthermal
  gas motions.
 While surprisingly similar in mass and size to the much better known Orion molecular cloud, the formation rate of high-mass
 stars appears to be suppressed in the California molecular cloud compared with that in the Orion molecular cloud based on the
 mass-radius threshold derived from the static Bonnor-Ebert sphere.
The largely uniform $^{12}$CO $J=2-1$ line-of-sight
 velocities along the L1482 molecular cloud shows that
 it is a generally coherent filamentary structure.
Since the NGC1579 stellar cluster is at the junction of two
 molecular filaments, the origin of the NGC1579 stellar
 cluster might be merging molecular filaments
  fed by converging inflows.
Our analysis suggests that these molecular filaments are
 thermally supercritical and molecular clumps may form by
 gravitational fragmentation along the filament.
Instead of being static, these molecular clumps are
 most likely in processes of dynamic evolution. }
{}
 \maketitle

\section{Introduction}

Recently, structures of interstellar filaments in molecular clouds
 (MCs) have been the subject of considerable research interest.
The {\it Herschel} Space Telescope with its $70-500\mu$m
 images taken in parallel mode by the SPIRE
 \citep{2010A&A...518L...3G} and PACS \citep{2010A&A...518L...2P}
 cameras on board reveals the omnipresence of parsec-scale
 molecular filaments in nearby MCs.
Filaments are detected both in unbound, non-star-forming complexes
 such as the Polaris translucent cloud
 \citep[e.g.][]{2010A&A...518L.103M,
 2010A&A...518L.104M, 2010A&A...518L..92W}  and in
 active star-forming regions such as the Aquila Rift cloud,
 where they are associated with the presence of prestellar
 clumps and protostars \citep[e.g.][]{2010A&A...518L.102A}.
These observational results suggest that molecular filament
 formation precedes any star-forming activities in MCs
 \citep[e.g.][]{2011A&A...529L...6A}.

Moreover, molecular filaments are promising sites to study
 the physics of MC formation and fragmentation as well as
 the earliest stages of forming protostars.
Filamentary structures appear to be easily produced by
 many numerical simulations of MC evolution that include
 hydrodynamic (HD) and/or magnetohydrodynamic (MHD) turbulence
 \citep[e.g.][]{2010A&A...512A..81F,2008A&A...486L..43H,2004RvMP...76..125M}.
On the other hand, theories and models for instabilities and
 fragmentation of the filament, or cylindrical gas structures
 have been extensively studied for five decades
 \citep[e.g.][]{1964ApJ...140.1056O,1997ApJ...480..681I}.
Therefore observations of molecular filaments are well-suited
 and necessary to test these theories and models.

The nearby California molecular cloud (CMC) has recently been
 recognized as a massive giant MC \citep[e.g.][]{2009ApJ...703...52L}.
The CMC is characterized by a filamentary structure and extends
 over about 10 deg in the plane of sky, which at a distance of
 $450\pm 23$ pc corresponds to a maximum physical dimension of
 $\sim 80\pm 4$ pc.
The molecular filament L1482 contains  most of the active
 star-forming
 regions in the CMC and  hosts the most massive young star
 Lk H$\alpha$ 101 \citep{1956PASP...68..353H}, which is a member
 of the embedded stellar cluster in NGC1579 and is likely an
 early B star \citep[e.g.][]{2004AJ....128.1233H}.
The molecular filament L1482 was reported for the first time as
 a dark nebula by \citet{1962ApJS....7....1L} five decades ago.
L1482 extends north and west along the filament by roughly
 1 degree \citep{2013ApJ...764..133H}.
Some earlier investigations of the CMC focus on large-scale structure
 with low resolutions of $^{12}$CO $J=1-0$  emissions, dust
 extinctions \citep[e.g.][]{2009ApJ...703...52L,2010A&A...512A..67L}
 , and dust continuum emissions \citep[e.g.][]{2013ApJ...764..133H}.
However, high-resolution molecular observations are very important
 and valuable for us to locally and globally understand the
  nature of the molecular filaments and the star formation
  activity inside the CMC.
Thus, L1482 offers an opportunity to study fragmentation of
 molecular filament and to investigate the kinematic and
 dynamical states of MCs and molecular clumps within the CMC.

In this paper, we  present a molecular  map
 study of the L1482 molecular filament in the  CMC.
 For the first time, observations of the molecular line
 transitions $^{12}$CO $J=2-1$, $^{12}$CO $J=3-2$ and
 $^{13}$CO $J=1-0$ along the entire molecular filament
  are displayed simultaneously.
Fragmentation of molecular filaments and dynamic properties
 of molecular clumps are probed and extensively discussed.

\section{Observations and database archives}
\subsection{KOSMA observations of molecular transition lines}

The  line emission maps of the $J=2-1$ and $J=3-2$ transitions
 of $^{12}$CO were made at the K\"olner Observatorium f\"ur
 Sub-Millimeter Astronomie
 (KOSMA) 3-meter telescope at Gornergrat Switzerland in March 2010.
The half-power beam widths of this telescope at the two observing
 frequencies are 130\arcsec at 230 GHz and 80\arcsec at 345 GHz.
The telescope pointing and tracking accuracy is
 better than 10\arcsec.
A dual channel SIS receiver for 230 GHz and 345 GHz was
 used for the frontend, with typical system temperatures
  120 K at 230 GHz and 150 K at 345 GHz.
The medium and variable resolution acousto-optical spectrometers
 with bandwidths 300 MHz at 230 GHz and $655-1100$ MHz at 345 GHz
  were used as the backends.
The spectral velocity resolutions were 0.22 km s$^{-1}$ at 230 GHz and
 0.29 km s$^{-1}$ at 345 GHz.
The 3m beam efficiencies B$_{\rm eff}$ are 0.68 at 230 GHz and 0.72
 at 345 GHz.
The forward efficiency F$_{\rm eff}$ is 0.93.
The mapping was made using the on-the-fly (OTF)
 mode with 1\arcmin$\times$ 1\arcmin grid size.
The correction for the molecular line intensities
 to the main-beam temperature scale was made using
 the formula T$_{\rm mb}$=F$_{\rm eff}$/B$_{\rm eff}
 \times$ T$_{\rm A}^{*}$.
Data reductions were carried out using the CLASS and GREG
 software packages, which are parts of GILDAS\footnote{%
  \tiny
GILDAS package was developed by the Institute de Radioastronomie
 Millim\'{e}trique (IRAM). http://www.iram.fr/IRAMFR/GILDAS.}.

\subsection{Archival data}

The $^{13}$CO $J=1-0$  map with a beam resolution of
 approximately 1 arcminute was retrieved from the
 Purple Mountain Observatory (PMO) archive \footnote{%
  \tiny
 Data acquired at website http://www.radioast.csdb.cn/.}.
  This map was obtained with the 13.7m radio telescope
 at Delingha Station in QingHai Province of west China.
The {\it Herschel}  spacecraft data were obtained
 by \citet{2013ApJ...764..133H} using the PACS detector
 at 70$\mu$m and 160$\mu$m bands \citep{2010A&A...518L...2P}
 and the SPIRE detector at 250$\mu$m, 350$\mu$m, and
 500$\mu$m bands \citep{2010A&A...518L...3G} onboard
 the {\it Herschel} Space Telescope.
The parallel scan mode of the {\it Herschel} Space
 Telescope was executed with the PACS at 70$\mu$m and 160$\mu$m
 for 5\arcsec and 12\arcsec resolutions \citep{2010A&A...518L...2P} and with the SPIRE at 250$\mu$m,
 350$\mu$m, and 500$\mu$m for 18\arcsec, 25\arcsec, and 36\arcsec resolutions \citep{2010A&A...518L...3G}) at a scan
 speed of 20\arcsec s$^{-1}$ for both the PACS and SPIRE cameras.
In this paper, we primarily focus on the NGC1579 stellar cluster
 and the L1482 filament regions as part of the CMC (see Fig.
 \ref{Fig01} below for more specific and detailed information).

\section{Results of the analysis}

\begin{figure}[h]
\vspace*{0.2mm}
\begin{center}
\includegraphics[width=9.2cm,trim=50 -15 -10 -10,clip]{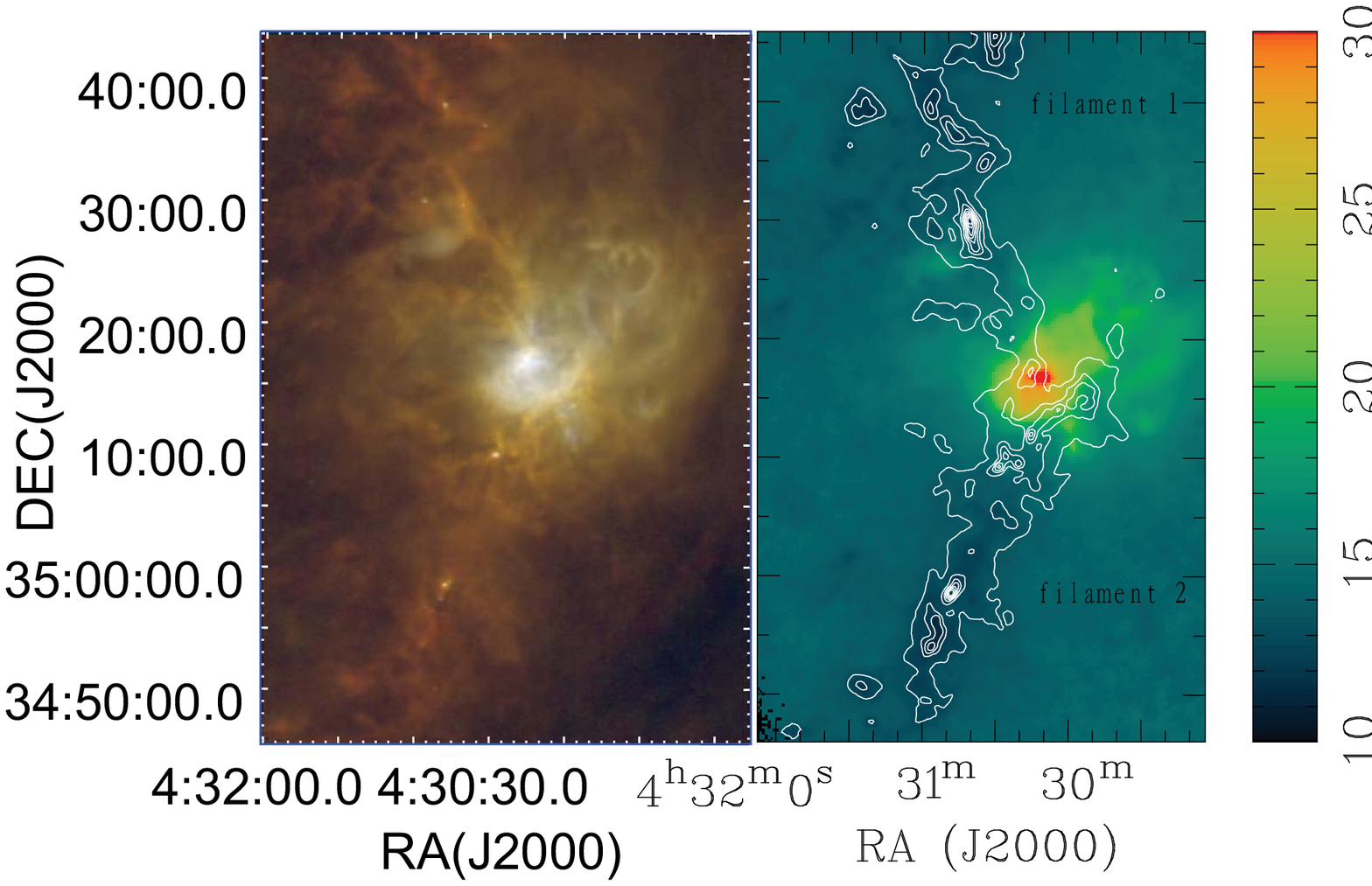}
\end{center}
\caption{Left panel: three-color image of molecular filament L1482;
  here, shows red the 250$\mu$m, green the 160$\mu$m, and
  blue the 70$\mu$m band \citep{2013ApJ...764..133H}.
The shorter wavelengths reveal hot dusts, such as the \ion{H}{ii}
  regions [Sharpless-222; see \citet{1959ApJS....4..257S}] around the NGC1579 stellar cluster, which are clearly shown in blue.
The longer wavelengths show the cold, dense objects, such as the
  cold dense filaments plotted in red .
Right panel: dust temperature map (36\arcsec corresponding
  to 0.08pc) of filament L1482, with the column density
  $N_{\rm H_2}$ of molecular hydrogen contours overlaid.
Contours are plotted from $6.3\times 10^{21}$ to
  $5.7\times 10^{22}$ cm$^{-2}$ with a spacing of
  $6.3\times 10^{21}$ cm$^{-2}$ (10\% to 90\% of the
  maximum spaced with 10\% contour interval).}\label{Fig01}
\end{figure}

 A three-color (red for $250 \mu$m, green for $160 \mu$m,
 and blue for $70 \mu$m) {\it Herschel} spacecraft image of the
  NGC1579 stellar cluster and L1482 molecular filament \citep{2013ApJ...764..133H} (left panel of Fig. \ref{Fig01}),
 consisting of prominent molecular filaments 1 and 2
 in the right panel of Fig. \ref{Fig01}, clearly
 shows relatively cold filamentary structures in the
 northern and southern parts of this ISM complex as
 well as fairly bright emissions at shorter wavelengths
 associated with the NGC1579 stellar cluster \citep[e.g.][]{1956PASP...68..353H,1971ApJ...169..537H}
  , which is also located at the junction of molecular
 filaments  (see Fig. \ref{Fig01}).
A few years ago, Schneider et al. (2012) studied the
 formation of stellar clusters in the Rosette molecular
 cloud and found all known infrared clusters except one
 lying at junctions of molecular filaments, as predicted
 by hydrodynamic turbulence simulations (e.g., Dale \&
 Bonnell 2011; Schneider et al. 2010).
These studies appear in general consistent with observed
 features of other ISM cloud complexes such as MCs
 Taurus, Ophiuchus, Rosette, and Orion
 \citep[e.g.][]{2009ApJ...700.1609M, 2012A&A...540L..11S, 2013ApJ...766L..17S}.


\subsection{Column density and temperature maps}

The column density $N_{\rm H_2}$ of molecular hydrogen and dust
 temperature maps of L1482 filament shown in the right panel
 of Fig. \ref{Fig01} are derived from \citet{2013ApJ...764..133H},
 who fit $160\mu$m, $250\mu$m, $350\mu$m, and $500\mu$m
 pixel-by-pixel for spectral energy distributions (SEDs) of
 \citet{2010A&A...518L.106K} at four longer {\it Herschel}
 wavebands (still with the same 36\arcsec spatial resolution)
 using a simplified model for dust emissions, namely
 $F_\nu=\kappa_\nu B(\nu,T)\times$(column density)
 with the assumed dust opacity law of
 $\kappa_\nu=0.1(\nu/1000{\rm\ GHz})^\eta {\rm\ cm}^2
 {\rm\ g}^{-1}$ and a spectral index $\eta=2$.
 Here, $B(\nu,T)$ is the Planck function defined by
 $B(\nu,T)=(2h\nu^3/c^2)[\exp(h\nu/kT)-1]^{-1}$,
 where $T$ denotes the absolute dust temperature,
 $\nu$ the frequency, $k$ the Boltzmann constant,
 $h$ the Planck constant, and $c$ the speed of light.


From the column density and temperature maps in the right
 panel of Fig. \ref{Fig01}, we note that the northern and southern
 parts of L1482 molecular filament are dominated by relatively cold
 dense gas materials, in contrast to the map center, where
 warmer dust is heated by the stellar cluster NGC1579.

  \begin{table*}
  \caption{Angular positions and the estimated physical parameters for the 23 identified molecular clumps.}
    \label{table:1}
     \centering
\begin{tabular}
{c c c c c c c}
\hline\hline
Source   & RA          & Decl      &  Peak Column Density               & T$_{\rm dust}$       &   Size           & Protostar$^\star$\\
 No. Label & (J2000)      & (J2000)   & (10$^{22}$ cm$^{-2}$)  &(K)              & [\arcsec$\times$\arcsec\ \ (\degr )] & Existence\\
\hline
 1    &  04:31:03.59  & +35:42:16.93    &  1.34      &  12.1    & 100 $\times$  77 ( 52.1 )       &  no \\
 2    &  04:30:56.90  & +35:39:45.37    &  1.85      &  11.5    & 234 $\times$ 103 (-90.0 )       &  no \\
 3    &  04:31:25.08  & +35:39:22.76    &  1.62      &  12.0    & 187 $\times$ 150 (-83.2 )       &  no \\
 4    &  04:30:47.05  & +35:37:27.99    &  2.18      &  12.0    & 251 $\times$ 103 ( 51.4 )       &  yes\\
 5    &  04:30:34.51  & +35:34:57.04    &  2.06      &  11.9    & 152 $\times$  86 (-38.4 )       &  no \\
 6    &  04:30:40.47  & +35:29:25.00    &  4.52      &  12.5    & 253 $\times$  89 ( 14.6 )       &  yes\\
 7    &  04:30:55.74  & +35:29:16.76    &  1.77      &  12.9    & 159 $\times$  97 ( 31.7 )       &  yes\\
 8    &  04:30:16.92  & +35:23:02.94    &  1.49      &  15.9    & 168 $\times$ 113 ( 13.7 )       &  no \\
 9    &  04:30:22.37  & +35:19:59.27    &  1.25      &  17.7    & 195 $\times$ 137 ( 56.2 )       &  no \\
10    &  04:30:16.15  & +35:17:06.51    &  1.83      &  29.7    & 228 $\times$ 110 (-22.3 )       &  yes\\
11    &  04:29:53.90  & +35:15:41.51    &  2.18      &  18.9    & 238 $\times$ 119 ( -5.0 )       &  yes \\
12    &  04:30:02.67  & +35:14:15.14    &  1.89      &  20.8    & 195 $\times$  68 (-63.9 )       &  yes \\
13    &  04:30:14.80  & +35:12:07.16    &  3.50      &  15.5    & 172 $\times$  85 (-19.1 )       &  no\\
14    &  04:30:26.60  & +35:10:44.14    &  1.55      &  14.2    & 119 $\times$ 108 (-14.8 )       &  no \\
15    &  04:30:20.06  & +35:09:38.44    &  2.32      &  13.3    & 119 $\times$  83 ( 60.4 )       &  no \\
16    &  04:30:29.88  & +35:08:54.14    &  1.87      &  13.6    & 124 $\times$ 100 (-74.6 )       &  no \\
17    &  04:30:13.94  & +35:07:46.97    &  1.80      &  13.4    & 103 $\times$  72 ( 28.8 )       &  yes\\
18    &  04:30:27.83  & +35:05:16.42    &  1.41      &  13.0    & 388 $\times$ 201 ( 44.1 )       &  no \\
19    &  04:30:38.95  & +35:02:12.12    &  1.27      &  12.5    & 293 $\times$ 189 (-27.9 )       &  yes\\
20    &  04:30:47.54  & +34:58:48.27    &  3.87      &  12.7    & 137 $\times$  68 (-33.6 )       &  yes\\
21    &  04:30:53.98  & +34:55:41.59    &  2.27      &  11.5    & 262 $\times$  93 ( 19.4 )       &  yes\\
22    &  04:30:56.40  & +34:54:11.01    &  2.62      &  11.1    & 115 $\times$ 102 (-32.6 )       &  yes\\
23    &  04:31:24.80  & +34:50:43.96    &  1.44      &  12.1    & 137 $\times$  89 ( 63.7 )       &  no \\
\hline 
 \end{tabular}
  \tablefoot{$^\star$ The protostar identifications  are taken from \citet{2009ApJS..184...18G}, \citet{2010ApJ...715..671W} and \citet{2013ApJ...764..133H}. }
   \end{table*}

\begin{table*}
\caption{Estimated physical parameters for the 23
  identified molecular clumps.}
\label{table:2}
\centering
\begin{tabular}
{c c c c c c c}
\hline\hline
Source       & Radius   & Mass    & n                & $\Delta v$          & $\sigma_{\mathrm{NT}}$       & f$_{\mathrm{turb}}$      \\
 No. Label   & (pc)     & ($M_\odot$)         & (10$^{4}$ cm$^{-3}$)     &(km $s^{-1}$)  &(km $s^{-1}$)     &   \\
\hline
  1  &  0.10   &   6.8    & 3.8      &  1.13           & 0.48         & 2.5        \\
 2   &  0.17   &  25.6    & 2.5      &  1.08           & 0.46         & 2.4          \\
 3   &  0.18   &  26.7    & 2.1      &                 &              &          \\
 4   &  0.18   &  32.7    & 2.9      &  1.29           & 0.54         & 2.9         \\
 5   &  0.12   &  15.8    & 3.9      &  1.44           & 0.61         & 3.2         \\
 6   &  0.16   &  61.3    & 6.8      &  1.47           & 0.62         & 3.2         \\
 7   &  0.14   &  16.4    & 3.2      &  1.20           & 0.51         & 2.6         \\
 8   &  0.15   &  16.7    & 2.4      &  1.36           & 0.57         & 2.6         \\
9    &  0.18   &  20.2    & 1.7      &  0.95           & 0.40         & 1.7         \\
10   &  0.17   &  25.4    & 2.4      &  1.25           & 0.52         & 1.7         \\
 11  &  0.18   &  37.5    & 2.9      &  1.65           & 0.70         & 2.9         \\
12   &  0.13   &  19.8    & 4.8      &  1.52           & 0.64         & 2.6         \\
13   &  0.13   &  18.9    & 4.0      &  1.47           & 0.62         & 2.9         \\
14   &  0.12   &  16.7    & 4.3      &  1.65           & 0.70         & 3.4         \\
15   &  0.11   &  16.2    & 6.2      &  1.72           & 0.73         & 3.6         \\
16   &  0.12   &  15.7    & 4.2      &  1.55           & 0.66         & 3.2         \\
17   &  0.09   &   6.8    & 4.0      &  1.52           & 0.64         & 3.2         \\
18   &  0.30   &  62.8    & 1.1      &  1.81           & 0.77         & 3.9         \\
19   &  0.26   &  43.3    & 1.2       & 1.32           & 0.56         & 2.9         \\
20   &  0.11   &  20.9    & 8.6      &  1.36           & 0.57         & 2.9         \\
21   &  0.17   &  31.1    & 3.0      &  1.24           & 0.52         & 2.8         \\
22   &  0.12   &  19.3    & 5.7                                                       \\
23   &  0.12   &  10.5    & 2.9    \\
\hline
\end{tabular}
  \end{table*}

\subsection{Molecular line emissions from molecular clouds}

\begin{figure}[h]
\vspace*{0.2mm}
\begin{center}
\includegraphics[width=11cm,trim=40 80 -30 85,clip]{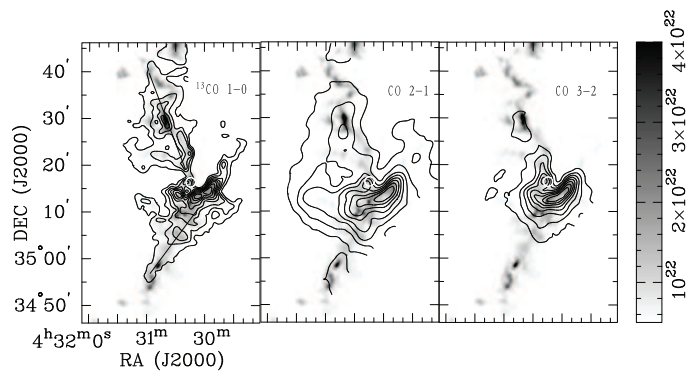}
\end{center}
\caption{Three molecular line transitions of
  $^{13}$CO $J=1-0$, $^{12}$CO $J=2-1$ and $^{12}$CO $J=3-2$
  shown as integrated molecular intensity maps are overlaid
  with the column density of hydrogen molecules
  (H$_2$).
The contours are from 20\% to 90\% of the peak
   with an adjacent contour interval of 10\%.
Molecular emissions integrated from $-3$ km s$^{-1}$ to
  3 km s$^{-1}$ (from the channel maps in Appendix A).
The
  white circle is the location of the stellar cluster
  NGC1579 and  the two straight lines in the left
  panel indicate the loci for the position-velocity
  diagrams in Fig. \ref{Fig07}.} \label{Fig02}
\end{figure}

\begin{figure}[h]
\vspace*{0.2mm}
\begin{center}
\includegraphics[width=6cm,trim=-15 5 0 0,clip]{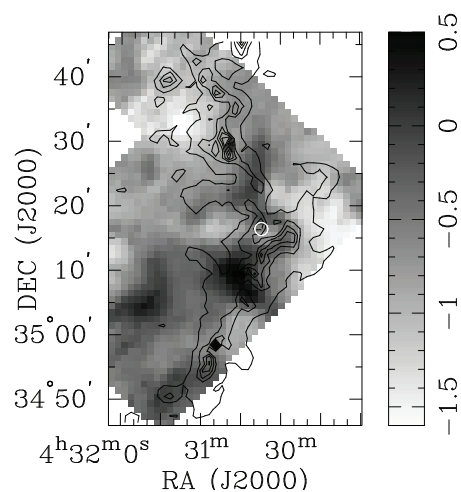}
\end{center}
\caption{Column density contours (black) are overlaid
   on the $^{12}$CO $J=2-1$ intensity-weighted
  velocity maps (first moment).
Contours here are plotted from  $6.3\times 10^{21}$ to
  $5.7\times 10^{22}$ with a spacing of  $6.3\times 10^{21}$
  (10\% to 90\% spaced with 10\% interval of the maximum).
The white
  circle  marks the location of the stellar
  cluster NGC1579.}\label{Fig03}
\end{figure}

The  various transitions of carbon monoxide CO spectral
 lines trace different molecular cloud environments.
For example, $^{13}$CO $J=1-0$ traces somewhat
 denser regions than $^{12}$CO $J=1-0$ does.
From Fig. \ref{Fig02} (the integrated molecular intensity maps
 overlaid by the $N_{\rm H_2}$ column density contours), we can
 see that the $^{13}$CO $J=1-0$ has a similar morphology as
 that of the dust column density, while $^{12}$CO $J=2-1$
  appears to trace a more extended component.
Since the upper energy level temperature and critical number
 density for $^{12}$CO $J=3-2$ line transition are 33.2 K
 and $5\times 10^4$ cm$^{-3}$, the molecular
 transition $^{12}$CO $J=3-2$ traces
 warmer and denser gas regions \citep[e.g.][]{1999ApJ...527..795K,2008A&A...484..361Q,2012A&A...544A..39J}.
 The $^{12}$CO $J=3-2$ line emission (Fig. \ref{Fig01}, right panel)
 suggests that the molecular gas could be heated by thermal
 emissions from dusts \citep[e.g.][]{1974ApJ...189..441G}.
The higher resolution velocity field (see Fig. \ref{Fig03} for the
 first-moment maps) from $^{12}$CO $J=2-1$ in gray scale is overlaid
 with the contours of the column densities of hydrogen molecules.
All molecular emissions are at the same line-of-sight velocity to
 within ($-1.5\sim 0.5$) km s$^{-1}$,  confirming that the
 filament is a single coherent object,  as proposed in \citet{2009ApJ...703...52L} and as first revealed by
 $^{12}$CO $J=1-0$ observations.

\subsection{Dense clumps and protostars embedded in filaments}
\subsubsection{Identifications of molecular clumps along filaments}

Based on the column density map, we used a two-step method
 \citep[e.g.][]{2007ApJ...655..351L} to identify molecular
 clumps in our acquired data and estimated their physical
 parameters.

First, we derived the root mean square (rms) of the column density
 map based on a background noise estimate of $1.7\times 10^{21}$
 cm$^{-2}$ from weak-emission regions, and then we used the
 Clumpfind2d algorithm \citep[e.g.][]{1994ApJ...428..693W}
 with a threshold of five times the rms noise level to identify
 the intensity peaks, which must be stronger than seven times
 the background rms noise level.
 Using the CLUMPFIND algorithm to locate emission-intensity
 peaks represents only the first step.

\begin{figure}[h]
\vspace*{0.2mm}
\begin{center}
\includegraphics[width=9cm,trim=-5 -10 -10 -5,clip]{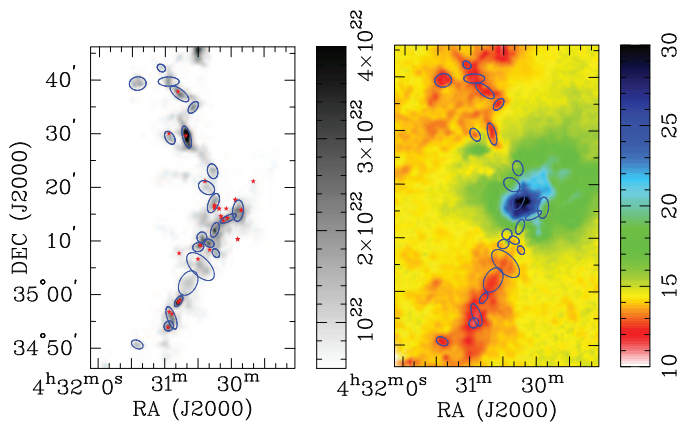}
\end{center}
\caption{Left panel: column density with dust clumps.
The red asterisk symbols represent protostars as determined by
  \citet{2009ApJS..184...18G}, \citet{2010ApJ...715..671W} and \citet{2013ApJ...764..133H}.
Right panel: dust temperatures overlaid with molecular hydrogen
  clumps marked by blue ellipses (in both panels).
  }
  \label{Fig04}
\end{figure}

Second,  we made a few assumptions about the boundary
 and shape of these identified molecular clump structures.
The boundary was defined as a 5 rms closed contour around the intensity
 peaks and the shape is a two-dimensional (2D) Gaussian structure.
We then used a 2D Gaussian procedure to fit identified molecular
 clumps within the 5 rms closed boundary around an intensity peak.
Finally, we also excluded isolated bright pixel spikes, which tend
 to appear towards edges of maps.
In summary,  by using this technique, we identified 23
 molecular clumps in  the column density map of L1482
 molecular filament with their basic properties
 summarized in Table \ref{table:1},  which includes,
 for each molecular clump, a numeric identification label,
 position, peak column density, and dust temperature.
In the last column of Table \ref{table:1}, we also indicate
 whether or not the identified molecular clump contains a
 protostar as determined previously in the
 literature \citep[e.g.]
 []{2009ApJS..184...18G,2010ApJ...715..671W,2013ApJ...764..133H}.
 Our simple coincidence criterion is that
 when the distance between the peak position of a molecular
 clump and a protostar is smaller than the semi-minor axis of
 the clump ellipse, the molecular clump is  regarded as
 containing a protostar.
 Positions of molecular clumps and  known protostars
 are  marked on both the column density and dust temperature
  maps in Fig. \ref{Fig04}.

The peak column densities of these molecular clumps fall in the
 range of [$1.25-4.52]\times 10^{22}$ cm$^{-2}$, with an average
 value of $ 2.08\pm 0.84\times 10^{22}$ cm$^{-2}$.
The peak dust temperature of these molecular clumps ranges from
 11.1 to 29.7 K, with an average value of  $14.4\pm 4.2$ K.


\subsubsection{Parameter estimates for the identified clumps }

 The masses of these identified molecular clumps are
 estimated by $M=\beta\ m_{\rm H_2} N_{\rm total,H_2}(D\Delta)^2$ \citep[e.g.][]{2012ApJ...761..171L,2012ApJ...760..147Q,2013ApJ...772..100L}, where $\beta=1.39$ converts the hydrogen mass into total mass taking into account the helium abundance  \citep[e.g.][]{1994ApJ...428..693W},
 $m_{\rm H_2}$ is the mass of
 a hydrogen molecule, ${N_{\rm total,H_2}}$ is the total estimated clump
 column density of H$_2$, $D\sim 450$ pc is the distance
 to the CMC, and $\Delta=14\arcsec$ is the pixel angular
 size of the data \citep[e.g.][]{2013ApJ...764..133H}.
The estimated molecular clump masses are shown in
 column 3 of Table 2.
The molecular clump masses range from 6.8 to $62.8M_\odot$,
 with an average value of $24.7_{-16.2}^{+31.1}M_\odot$.
The average mass of molecular clumps containing
 protostars is $28.6M_\odot$, while the average
 mass of the starless clumps is $21.0M_\odot$.
Thus,  there might be a  slightly increased
 average mass in the case of molecular clumps with embedded
  protostars than in that of starless
 molecular clumps, although it
 may not represent a significant systematic difference.
The radii of the identified molecular clumps are defined
 by $R=\sqrt{ab}/2$, where $a$ and $b$ are
 the major and minor axes of a clump ellipse.
 The estimated clump radii are listed
 in column 2 of Table 2 and range from 0.10 to 0.30 pc, with
 an average value of $0.15\pm 0.05$ pc.
The volume number densities for molecular hydrogen
 H$_2$ of these clumps are inferred according to the
 expression $n=3M/(4\pi R^3 m_{\rm H_2})$.
The estimated volume number densities are shown
 in column 4 of Table 2, they range from 1.1 to
 $8.6\times 10^4$ cm$^{-3}$, with an average
 value of $3.7\pm 1.8\times 10^4$ cm$^{-3}$.
 The average radius $\sim 0.15$ pc of the
 identified molecular clumps in our data analysis
 is comparable to that of the identified CS clumps
 in the molecular cloud Orion A, $\sim 0.16$ pc, and that
 of the identified NH$_3$ clumps in the molecular cloud
 Perseus, $\sim 0.12$ pc. But mean mass of the identified molecular clumps
 of our data analysis is considerably smaller than that of the CS
 clumps in the molecular cloud Orion A, $> 100M_\odot$
 \citep[e.g.,][]{1993ApJ...404..643T}, and is systematically
 larger than that of the NH$_3$ clumps in the molecular cloud Perseus,
 $\sim 9M_\odot$ \citep[e.g.,][]{1994ApJ...433..117L}.

\citet{2010ApJ...723L...7K} gave an empirical mass-radius
 threshold for massive star formation, namely
 $M(r)=870 M_\odot(r/\rm pc)^{1.33}$, above which massive
 stars form.
We invoke this criterion to examine our identified molecular
 clumps $-$ their masses as a function of radius are shown in
 Fig. \ref{Fig06}.
Based on this empirical mass-radius threshold, our identified
 molecular clumps may not form massive stars.
\citet{2013ApJ...768L...5L} found $\sim 40$\% of
 the molecular clumps in the Orion nebula above
 this threshold.
While strikingly similar in mass and size to the better known Orion molecular cloud (OMC), the formation
 rate for high-mass stars appears to be more depressed in the
 CMC than that in the OMC.


\section{Analysis and discussion}
\subsection{Gravitational instabilities
 for dense molecular clumps}
\subsubsection{Velocity dispersion and turbulence}

\begin{figure}[h]
\vspace*{0.2mm}
\begin{center}
\includegraphics[width=8.7cm,trim=55 12 20 10 ,clip]{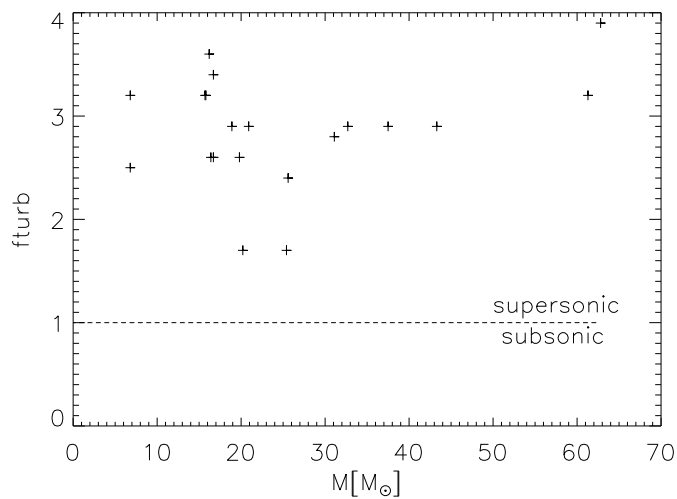}
\end{center}
\caption{
  $f_{\mathrm{turb}}=\sigma_{\mathrm{NT}}/\sigma_{\mathrm{Therm}}$
  ratios versus molecular clump masses in units of the solar mass
  $M_\odot$.
The horizontal dashed line for
  $\sigma_{\mathrm{NT}}/\sigma_{\mathrm{Therm}}=1$ is the
  demarcation between subsonic and supersonic nonthermal motions. }
  \label{Fig05}
\end{figure}

We used the molecular transition $^{13}$CO $J=1-0$ linewidths to
 calculate the nonthermal contribution of the line-of-sight
 velocity dispersion (averaged over one 1\arcmin\ beam size)
 and the level of internal flow turbulence.
The nonthermal ($\sigma_{\mathrm{NT}}$) and thermal
 ($\sigma_{\mathrm{Therm}}$) one-dimensional velocity
 dispersions in  our identified molecular clumps
 were calculated  according to the formulae of
 \citet{1983ApJ...270..105M}:
 \begin{equation}
       \sigma_{\mathrm{NT}} =\bigg(\sigma_{^{13}\rm CO}^2- \frac{kT_{\rm kin}}{m_{^{13}\rm CO}}\bigg)^{1/2}  \,, \label{eq01}
   \end{equation}
 \begin{equation}
       \sigma_{\mathrm{Therm}}
       =\bigg(\frac{kT_{\rm kin}}{m_{\rm H}\mu}\bigg)^{1/2} \,, \label{eq02}
   \end{equation}
where $\sigma_{^{13}\rm CO}$ is the one-dimensional velocity
 dispersion of $^{13}$CO$(1-0)$ related to the FWHM
 linewidth ($0.95\sim 1.81$ km s$^{-1}$ with an
 average of $1.40\pm 0.22$ km s$^{-1}$; see column 5
 of Table \ref{table:2}) as $\sigma=\Delta v/2.355$.
Here, the dust temperature is taken as T$_{\rm kin}$ with
 $k$ being the Boltzmann constant, $m_{^{13}\rm CO}$ is
 the mass of a $^{13}$CO molecule, $m_{\rm H}$ is the mass
 of an atomic hydrogen, and $\mu=2.72$ is the mean molecular
 weight of  a molecular hydrogen ISM gas.
Furthermore, the level of internal flow turbulence
 is given by the ratio $f_{\mathrm{turb}}= \sigma_{\mathrm{NT}}/\sigma_{\mathrm{Therm}}$, where $\sigma_{\mathrm{Therm}}$ is the one-dimensional
 isothermal sound speed  in a molecular hydrogen
 ISM cloud.
The $f_{\mathrm{turb}}=\sigma_{\mathrm{NT}}
 /\sigma_{\mathrm{Therm}}$ ratios versus the molecular
 clump masses are shown in Fig. \ref{Fig05} and all
 molecular clumps appear to be supersonic.
The values of $\sigma_{\mathrm{NT}}$ and
 $f_{\mathrm{turb}}$ are given in columns 6 and 7
 in Table \ref{table:2}.
The $\sigma_{\mathrm{NT}}$ ranges from 0.40 to 0.77
 km s$^{-1}$ with an average value of $0.59\pm 0.10$
 km s$^{-1}$.
The $f_{\mathrm{turb}}$ ranges from 1.7 to 3.9
 with an average value of $2.9\pm 0.5$.

\subsubsection{Static Bonnor-Ebert sphere model}

To examine the dynamic state of molecular clumps, we assumed that these
 clumps are only thermally supported against the gas self-gravity.
\begin{figure}[h]
\vspace*{0.2mm}
\begin{center}
\includegraphics[width=9cm,trim=20 12 0 10 ,clip]{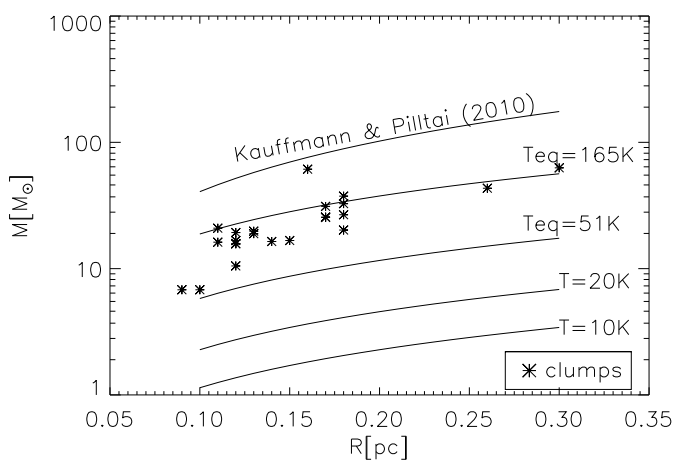}
\end{center}
\caption{
Mass (in units of solar mass $M_\odot$) versus radius
  (in units of pc) of the molecular clumps in our sample
  compared with the static Bonnor-Ebert sphere.
The top curve [e.g. Kauffmann \& Pilltai (2010)] represents
  the mass-radius threshold for forming massive stars.
The other curves are the critical masses that can be supported
  for clumps of specified temperatures and sizes.
  T$_{\mathrm{eq}}=51$ K corresponds to a turbulent linewidth
  of $\sim 1$ km s$^{-1}$ (FWHM), and 165 K corresponds to a
  linewidth of $\sim 1.8$ km s$^{-1}$.
   }
  \label{Fig06}
\end{figure}
We computed the static Bonnor-Ebert mass-radius relation for a
 spherical, self-gravitating, isothermal, and hydrostatic
 molecular clump.
The Bonnor-Ebert mass is obtained by integrating the mass density
 profile of the Bonnor-Ebert sphere, with an approximate analytic
 expression invoked by \citet{2004A&A...416..191T} as follows:
 \begin{equation}
      \rho(\xi)=1/[1+(\xi/2.25)^{2.5}]\ \,, \label{eq03}
   \end{equation}
\begin{equation}
      \xi = r(4\pi G\rho_c/v_s^2)^{1/2}\ \,, \label{eq04}
   \end{equation}
where $\rho_c$ is the central mass density and
 $v_s=[kT/(\mu m_H)]^{1/2}$
 is the isothermal sound speed.
\citet{1955ZA.....37..217E} and \citet{1956MNRAS.116..351B}
 pointed out that when the dimensionless radial variable
 $\xi$ exceeds a critical value of 6.5, there is no longer
 any stable Bonnor-Ebert sphere solution.
Assuming $\xi_{max}=6.5$ and using the observed clump radius,
 the central mass density $\rho_c$ can be determined and
 then the critical clump mass $M_{\rm BE-CR}$ can be readily
 calculated by integrating equations (\ref{eq03}) and (\ref{eq04}).

In Fig. \ref{Fig06}, the sizes and masses of molecular clumps are
 plotted along with the calculated curves based on critical static
 Bonnor-Ebert (BE) spheres at various gas temperatures.
Our identified molecular clumps appear to be too massive
 to be stable BE spheres for a kinetic temperature as
 high as $\sim 20$ K.
One important caveat for this analysis and discussion is
 that the observed size of our molecular clumps are not
 as clearly defined as required by the theoretical model.
Nevertheless, the considerable excesses of the observed
 molecular clump mass relative to the BE critical mass do
 suggest that increasing the clump size by as much as a
 factor of a few would not make a clump thermally stable
 against the self-gravity.
The important significance of our data analysis
 is that our identified molecular clumps here
 are not expected to be in static equilibria.
These molecular dense clumps are most likely in
 dynamic processes of being formed on the spot
 \citep{2004MNRAS.348..717L,2004ApJ...611L.117S,2006MNRAS.370L..85S}.
It seems apparent that dynamic evolutions of these star-forming
 and/or dense starless molecular clumps are inevitable
 \citep[][]{2006MNRAS.373.1610L,2009MNRAS.400..887G,2010MNRAS.403.1919G}.
For quasi-spherical molecular clumps in hydrodynamic evolution,
 it is possible to use at least four major observational
 diagnostics to constrain self-consistent dynamic models
 coupled with radiative transfer calculations \citep[e.g.][]{2011MNRAS.412.1755L,2011ApJ...741..113F},
 namely, extensive observations of various properties of
 molecular lines (e.g. line profiles), submillimeter
 continuum emissions from thermal dusts, and dust extinction
 observations as well as neutral hydrogen 21cm absorption
 lines in MCs.

To evaluate the importance of gas flow turbulence
 in molecular clumps, we also considered a modified
 BE sphere model taking into account turbulence
 in an empirical manner \citep{2003AJ....126..311L}.
We can define an equivalent temperature as
\begin{equation}
 T_{\mathrm{eq}}=\frac{m_{\mathrm{H}} \Delta V^2}{8 \ln (2) k} \,,
 \end{equation}
where $\Delta V$ is the total FWHM linewidth of the gas.
The observed FWHM linewidth of $^{13}$CO $J = 1-0$ at $1\arcmin$
 scale towards these regions ranges from 1.0 to 1.8 km s$^{-1}$.
We have plotted the critical mass curves based on $T_{\rm eq}$
 ($\Delta V=1$ km s$^{-1}$ ) and $T_{\rm eq}$ ($\Delta V=1.8
 $km s$^{-1}$) in Fig. \ref{Fig06}.
Assuming a FWHM linewidth of 1 km s$^{-1}$, most of molecular
 clumps appear to be supercritical.
This is in contrast to  molecular clumps identified in
  the Taurus molecular cloud, where  most molecular
 clumps remain subcritical \citep[e.g.][]{2012ApJ...760..147Q}.

\subsection{ Fragmentation along a molecular filament}
\subsubsection{ Formation of molecular clumps}

The formation of molecular clumps may proceed in two main
 stages based on {\it Herschel} observational results \citep[e.g.][]{2010A&A...518L.102A,2011IAUS..270..255A}.
First, large-scale magnetohydrodynamic (MHD) turbulence generates
 a complex network of molecular filaments in the interstellar medium
 (ISM) \citep[e.g.][]{2001ApJ...553..227P,2001MNRAS.327..715B}.
In the second stage, the self-gravity takes over and fragments
 the densest molecular filaments into prestellar clumps via
 gravitational collapse instabilities \citep[e.g.][]{1997ApJ...480..681I,1964ApJ...140.1056O}.

The critical mass per unit length, $M_{line}^{crit}=2c_s^2/G$ (where $c_s$ is the isothermal sound speed and $G$ is the
 gravitational constant) is the critical value required for a
 molecular filament to be gravitationally unstable to radial
 contraction and fragmentation along its length \citep[e.g.][]{1997ApJ...480..681I}.
Remarkably, this critical line mass only depends on the
 gas temperature \citep{1964ApJ...140.1056O}.
It is approximately equal to $\sim$ 16 M$_\odot$/pc for
 molecular gas filaments at T$=10$ K, corresponding to
 a $c_s\sim 0.2$ km s$^{-1}$.
\citet{2013A&A...553A.119A} assumed that ISM filaments
 have Gaussian radial column density profiles,
 an estimate of the mass per unit lengh is given by M$_{line}$
 $\approx\Sigma_0\times W_{fil}$, where $W_{fil}$ is the typical
 filament width \citep{2010A&A...518L.102A,2013A&A...553A.119A}
 and $\Sigma_0=\mu\ m_{\rm H} N_{\rm H_2^0}$ is the central gas
 surface mass density of a filament.
For a typical filament width of $\sim$ 0.1 pc, the theoretical
 value of $\sim$ 16 M$_\odot$/pc corresponds to a central column
 density $N_{\rm H_2}^0\sim 8\times 10^{21}$ cm$^{-2}$.
These filaments may be grouped into thermally supercritical and
 thermally subcritical filaments depending on whether their line
 masses higher or lower than 16 M$_\odot$/pc correspond to a
 central column density $N_{\rm H_2}^0\sim 8\times 10^{21}$ cm$^{-2}$.
Thermally supercritical filaments are expected to be globally
 unstable to radial gravitational collapse and fragmentation
 into prestellar clumps along their lengths.
The prestellar clumps formed in this manner are themselves
 expected to collapse locally into protostars in dense
 molecular clumps.

Since the central number density of molecular filament 1
 ( the average value of peak density for all clumps
 located within molecular filament 1) is $\sim 2.0\times
 10^{22}$ cm$^{-2}$ and that of molecular filament 2
 ( the average value of the peak density for all clumps located
 within molecular filament 2) is $\sim 2.3\times 10^{22}$ cm$^{-2}$
 (both above $8\times 10^{21}$ cm$^{-2}$), they all appear to be
 thermally supercritical filaments.
 Therefore, the L1482  filamentary cloud appears
 to be a thermally supercritical  structure, susceptible
 to gravitational fragmentation.
 \citet{2013MNRAS.436.3707L}
 studied magnetic fields of filamentary clouds in the Gould
 Belt and found that the density threshold of cloud gravitational
 contraction is $\sim 10^{22}$ cm$^{-2}$.
Still, these molecular filaments remain susceptible to gravitational
 fragmentation, in qualitative agreement with the detected clumps
 and protostars.
This last stage of filament-to-clumps evolution seems to emerge
 as a phenomenon often found in recent observational studies.

We have already indicated in the earlier discussion that MHD
 turbulence is likely to play a major role in creating the
 filamentary network in the magnetized ISM according to
 observations and numerical simulations.
For an individual magnetized filament with a mean magnetic
 field $B_0$ along the filament, the Jeans length for
 gravitational collapse instability transverse to the filament
 $\lambda_{\perp}=[\pi (C_S^2+C_A^2)/(\rho_0 G)]^{1/2}$
 becomes larger \citep{1953ApJ...118..116C,1996MNRAS.279L..67L},
 where $C_S$ is the sound speed and
 $C_A=B_0/(4\pi\rho_0)^{1/2}$ is the Alfv\'en wave speed.
Although the Jeans length for gravitational
 collapse instability along the filament remains the
 same as the hydrodynamic one, the growth rates of these
 instabilities change and become anisotropic due to the
 very presence of the mean magnetic field $B_0$.
For a large-amplitude circularly polarized incompressible
 Alfv\'en wave train propagating along such a mean magnetic
 field aligned with an ISM filament, the Jeans lengths
 for gravitational collapse instabilities are increased
 in all directions anisotropically because a large-amplitude
 Alfv\'en wave train involves a transverse magnetic field
 relative to the mean $B_0$ direction.
Therefore, such magnetized collapsed clumps should have
 systematically larger masses, depending on the magnetic
 field strengths \citep{1996MNRAS.279L..67L}.
For a magnetized ISM with MHD turbulence, both turbulent flow
 velocities and random magnetic fields tend to increase the
 volume and mass of a collapsed molecular clump in general \citep[e.g.][]{1966MNRAS.132..359S,1987QJRAS..28..197R}.
It is therefore of considerable interest to measure
 magnetic field strengths and turbulent flow velocities
 in such ISM filamentary network and dense molecular
 clumps embedded in individual filaments.
 Without any polarimetric observations available at the
 moment, the presence of a strong, permeating magnetic field
 in MCs and molecular filaments remains still hypothetical.

Empirically, magnetic field strengths in the ISM may
 range from a few $\mu$G to several hundred $\mu$G.
By the common wisdom of equipartition in energies,
 we tend to suggest that the sound speed $C_S$
 and the Alfv\'en speed $C_A$ may be comparable.
Along this line of arguments for an ISM with random
 magnetic fields, the Jeans volume for a collapsing
 molecular clump may be $\sim$ three times larger than
 that in the absence of magnetic field.
By these estimates, the molecular clumps studied here
 remain supercritical even when an equipartition random
 magnetic field is taken into account.

\subsubsection{ Role of turbulence in molecular clumps}

\begin{figure}[h]
\vspace*{0.2mm}
\begin{center}
\includegraphics[width=8cm,trim=0 0 300 0,clip]{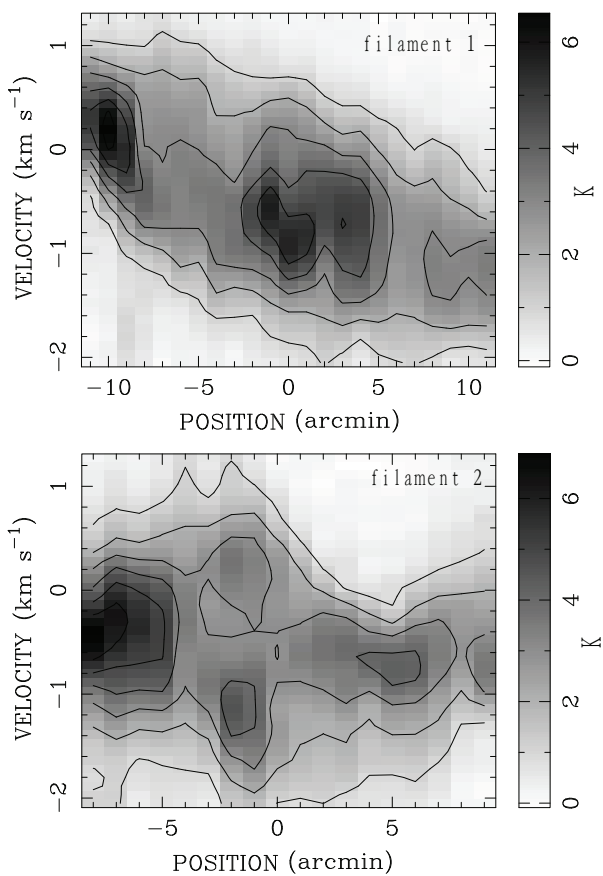}
\end{center}
\caption{Position-velocity diagrams of the $^{13}$CO
  $J=1-0$ emission along the two marked straight lines
  in the left panel of Fig. \ref{Fig02}.
The contour levels in both panels are 10\% to 60\% of the
  peak value spaced with 10\% contour interval, successively.
  }\label{Fig07}
\end{figure}

 The $^{13}$CO $J=1-0$ line emission probes
 nonthermal supersonic linewidths, implying that the
 internal flow structures of a $\sim$ MC, molecular clumps
 and their environments may be highly turbulent.
As such, (MHD) turbulence would play an important
 role in the process of dynamic fragmentation.
 It might be the case that CO isotopes are tracing a
 low-density, more turbulent component of the gas,
 that these clumps are most likely substructured as
 indicated by turbulent fragmentation models, and that
 with higher resolution, a different dense gas tracer
 is required to resolve individual cores of smaller scales with more coherent
 gas motions.
In particular, interferometric observations with a tracer
 [e.g. CS(2-1)] of
 higher resolution are necessary to demonstrate that further
 fragmentation within each clump (broken into groups of $4-5$
 pieces) may occur at the prestellar stage, as in the case
 of the OMC \citep[e.g.][]{2013ApJ...772..100L}.
Approximately along the central axes of molecular filaments
 1 and 2 as shown in Fig. \ref{Fig02}, we made their
 position-velocity diagrams in Fig. \ref{Fig07}.
Filament 1 clearly shows velocity gradients.
These gradients might be related to inflows along
 the filament \citep[e.g.][]{2013ApJ...772..100L}.
For example, molecular filament 1 might involve
 supersonic converging flows along the filament
 to feed embedded molecular clumps containing
 protostars and/or cluster of subcomponents.

\section{Conclusion and summary}

We have conducted a comparative and comprehensive mapping
  study for molecular transition lines of $^{13}$CO $J=1-0$
  (Delingha), $^{12}$CO $J=2-1$ and $^{12}$CO $J=3-2$ (KOSMA)
  and multiband infrared data of {\it Herschel} spacecraft
  observations towards the molecular filament L1482 in CMC.
The main results of our data analysis and parameter
  estimates are summarized below.

(1) The L1482 molecular filament is most likely a spatially coherent
 global structure, as shown by the more or less uniform
 $^{12}$CO $J=2-1$ line-of-sight velocities throughout the MC.

(2) As the NGC1579 stellar cluster is physically located at the
 junction of the two main molecular filaments 1 and 2, we propose that NGC1579 stellar cluster might originate from
 molecular filament merger processes involving converging inflows
 to feed the cluster and molecular clumps.

(3) We have identified 23 molecular clumps in this region.
  The peak column densities of these molecular clumps
 range from 1.25 to $4.52\times 10^{22}$ cm$^{-2}$, with
 an average value of $2.08\pm 0.84\times 10^{22}$ cm$^{-2}$.
The peak dust temperature of these molecular clumps ranges
 from 11.1 to 29.7 K, with an average value of $14.4\pm 4.2$ K.
The radii of these molecular clumps range from 0.10 to
 0.30 pc, with an average value of $0.15\pm 0.05$ pc.
The number densities range from 1.1 to $8.6\times10^4$ cm$^{-3}$,
 with an average value of $3.7\pm 1.8\times 10^4$ cm$^{-3}$.
The masses range from 6.8 to 62.8$M_\odot$, with an average
 value of $24.7_{-16.2}^{+31.1}M_\odot$.
Eleven molecular clumps are found to be associated with protostars,
 and the star-formation processes are most likely ongoing.
If follows that the defined protostellar and starless subsamples
 of molecular clumps have similar temperatures and linewidths,
 but  perhaps different masses, with an average mass of
 $28.6 M_\odot$ for the
 former, which is slightly more massive than $21.0 M_\odot$
 than the latter;  it would be of considerable interests to
 further confirm whether this is indeed a systematic difference,
 because resolving substructures of molecular clumps is impossible
 with the currently available data.
 The average radius of the identified molecular clumps in
 our data analysis, $\sim 0.15$ pc,
 is comparable to that of the identified CS clumps in the
 molecular cloud Orion A, $\sim 0.16$ pc, and that of the
 identified NH$_3$ clumps in the molecular cloud Perseus, $\sim 0.12$ pc.
However, the mean mass of the identified molecular clumps in
 our data analysis is considerably smaller than that of the
 identified CS clumps in the molecular cloud Orion A, $>100M_\odot$,
 and is systematically larger than that of the identified
 NH$_3$ clumps in the molecular cloud Perseus, $\sim 9M_\odot$.


(4) The $\sigma_{\rm NT}$ of these molecular
 clumps ranges from 0.40 to 0.77 km s$^{-1}$,
 with an average value of $0.59\pm 0.10$ km s$^{-1}$.
The ratio $f_{\rm turb}$ of these molecular clumps ranges from
 1.7 to 3.9, with an average value of $2.9\pm 0.5$.
Our identified molecular clumps typically show supersonic
 nonthermal motions according to the estimates.

(5) Based on the empirical mass-radius threshold  of
 a static Bonnor-Ebert sphere model with effects of turbulence
 empirically included via an effective temperature
 (e.g. Lai et al. 2003),
  our identified molecular clumps may not be able to
  form massive stars.
While surprisingly similar in mass and size to the well-known
 Orion molecular cloud, the rate of forming high-mass stars
 is considerably lower in the CMC than that
 in the OMC.

(6) Our critical mass per unit length estimates suggest that
 molecular filaments examined here are thermally supercritical
 and molecular clumps most likely involve dynamic processes
 of gravitational fragmentation along the molecular filaments.

(7) By the equipartition argument for energies, the presence
 of mean as well as random magnetic fields tends to increase
 the Jeans mass of a collapsing mass.
The masses of our identified molecular clumps appear to be still supercritical even in the presence of such
 equipartition magnetic fields.
This seems to suggest that instead of being static,
 these molecular clumps are most likely in the stage of
 hydrodynamic or MHD evolution.

 (8) Molecular filament 1 shows clear velocity gradients.
These gradients might be partly associated with inflows along
 filaments.
Such flows may play important roles of feeding stellar cluster
 and molecular clumps during star formation processes along
 molecular filament L1482.

\begin{acknowledgements}

We thank the anonymous referee for helpful comments
 and suggestions to improve the quality of the manuscript.
We also thank T. Liu  and others for providing
 help with the data reduction and analysis.
This research work was funded by the National Natural
 Science Foundation of China (NSFC) under grant 10778703
 and partially supported by
the National Basic Research Program of China (973 program,
 2012CB821802) and the National Natural
 Science Foundation of China (NSFC) under grants 11373062, 11303081,
 10873025, 10373009, 10533020, 11073014 and J0630317, and the MoE
 SRFDP 20050003088, 200800030071 and 20110002110008.
\end{acknowledgements} %

\bibliographystyle{aa} 
\bibliography{l1} 

\Online
\onecolumn
\begin{appendix} 

\section{$^{12}$CO $J=2-1$ and $^{12}$CO $J=3-2$
 channel maps as derived from our KOSMA observations
 in the Swiss Alps.}

\begin{figure}[h]
\vspace*{0.2mm}
\begin{center}
\includegraphics[width=9cm,angle=-90]{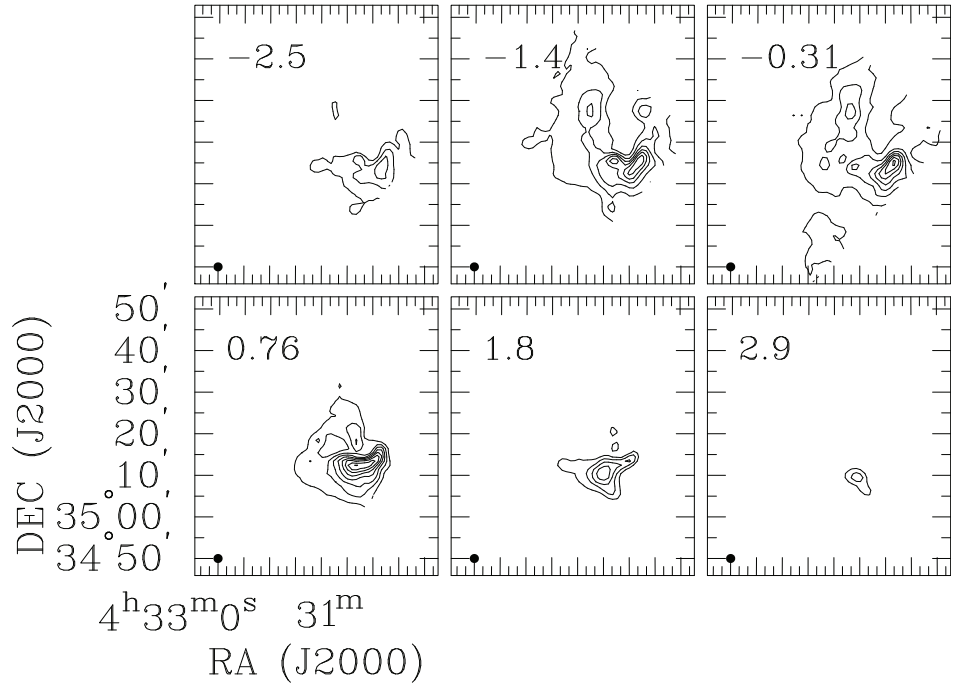}
\end{center}
\caption{Molecular transition $^{12}$CO $J=2-1$ channel maps.
The center
  velocities are shown in the upper left
  corners of each panel in units of km s$^{-1}$.
The integrated velocity interval in each panel is 1.1 km s$^{-1}$.
The contour levels are from 20\% up to 90\% of the peak with an
  adjacent contour intervals of 10\%.
The black solid circle in the lower left corner gives the beam size. }
  \label{Fig08}
\end{figure}

\begin{figure}[h]
\vspace*{0.2mm}
\begin{center}
\includegraphics[width=9cm,angle=-90]{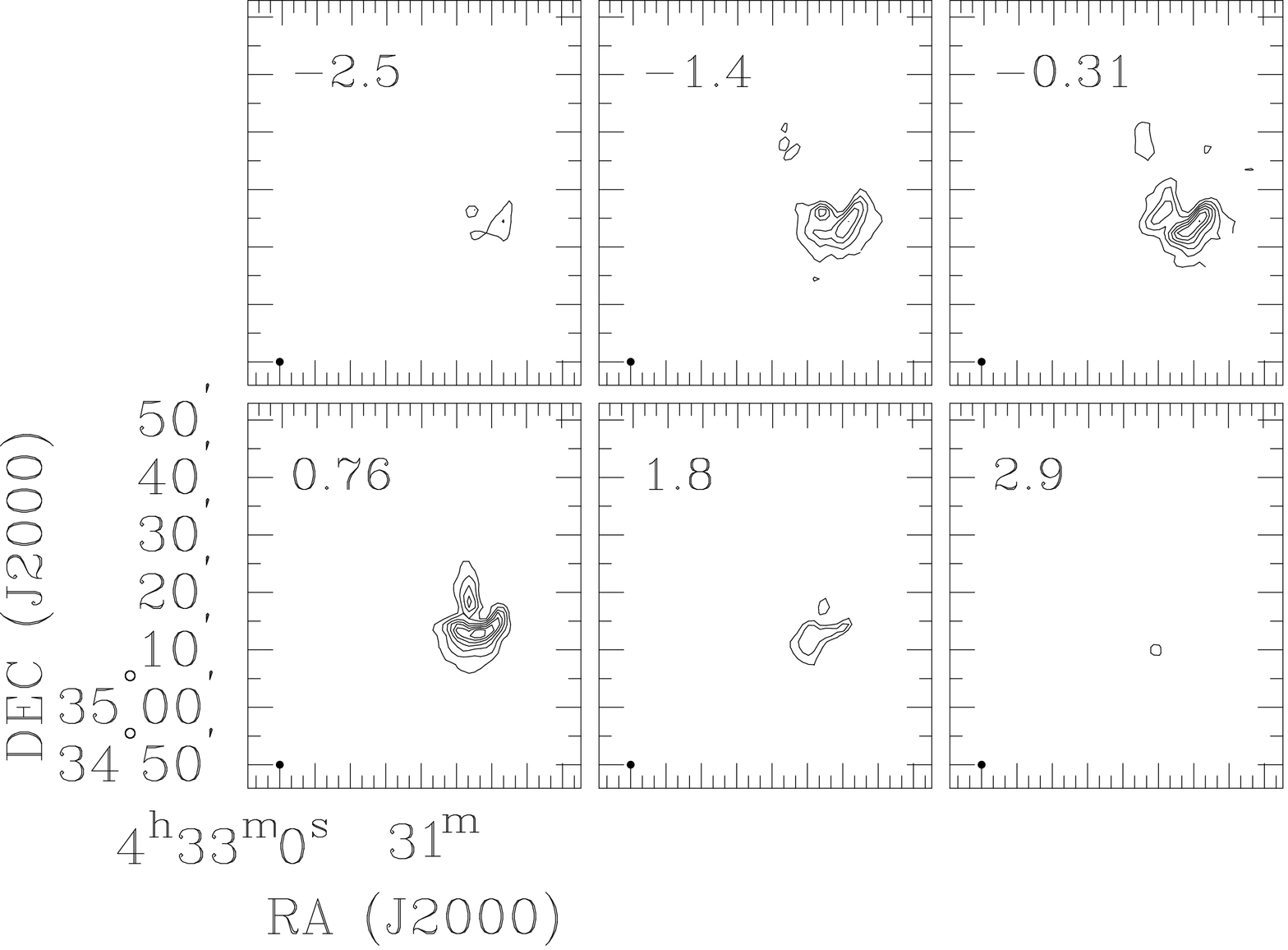}
\end{center}
\caption{Molecular $^{12}$CO $J=3-2$ channel maps.
The center
  velocities are shown in the upper
  left corners of each panel in units of km s$^{-1}$.
The integrated velocity interval in each panel is 1.1 km s$^{-1}$.
The contour levels are from 20\% up to 90\% of the peak with
  an adjacent 10\% contour interval.
The black solid circle in the lower left corner gives the beam
  size.}\label{Fig09}
\end{figure}

Figs. \ref{Fig08} and \ref{Fig09}
 present channel maps for the two molecular
  line transitions $^{12}$CO $J=2-1$ and $^{12}$CO $J=3-2$
  integrated over narrow ranges in $V_{\rm LSR}$.
The molecular line data were acquired by the KOSMA 3-m
  telescope in the Swiss Alps.
Detailed descriptions of the observations can be found in Section 2.

\end{appendix}

\end{document}